\documentclass[10pt,aps,pra,twocolumn,preprintnumbers,amsmath,amssymb,nofootinbib,superscriptaddress]{revtex4-2}
\usepackage{graphicx,longtable,mathrsfs,color,array}
\usepackage{hyperref}
\usepackage[usenames,dvipsnames]{xcolor} 
\usepackage{amssymb,amsmath,mathtools,mathrsfs,slashed} 
\usepackage{epsfig,subfigure,placeins,float} 
\usepackage{booktabs,longtable,ctable,multirow} 
\usepackage{exscale,relsize} 
\usepackage[normalem]{ulem} 
\usepackage{enumerate}
\usepackage{enumitem}
\usepackage{comment}
\usepackage{bm}
\usepackage{color}
\usepackage{braket}
\allowdisplaybreaks[1]

\newcommand{\be}{\begin{equation}}
\newcommand{\ee}{\end{equation}}
\newcommand{\bea}{\begin{eqnarray}}
\newcommand{\eea}{\end{eqnarray}}

\newcommand{\kvec}{{\bm k}}
\newcommand{\xvec}{{\bm x}}
\newcommand{\early}{{\mbox{\tiny early}}}
\newcommand{\late}{{\mbox{\tiny late}}}
\newcommand{\mt}{$\tilde{m}$}
\newcommand{\kt}{$\tilde{k}$}

\begin{document}

\title{Resonant Gravitational Production of Chiral Dark Photons}

\author{Leah Jenks}
\email{ljenks3@jh.edu}
\affiliation{William H.\ Miller III Department of Physics \& Astronomy,\\ Johns Hopkins University, 3400 N.\ Charles St., Baltimore, MD 21218, USA}

\author{Marc Kamionkowski}
\email{kamion@jhu.edu}
\affiliation{William H.\ Miller III Department of Physics \& Astronomy,\\ Johns Hopkins University, 3400 N.\ Charles St., Baltimore, MD 21218, USA}

\author{Edward W. Kolb}
\email{ekolb@uchicago.edu}
\affiliation{Kavli Institute for Cosmological Physics and Enrico Fermi Institute,\\ The University of Chicago, 5640 S.\ Ellis Ave., Chicago, IL, 60637, USA}

\begin{abstract}

Vector fields with a chiral, parity-odd coupling to a pseudoscalar field are known to undergo resonant amplification. In this work, we study this phenomenon in the context of cosmological gravitational particle production during early-universe inflation. We consider a massive vector (dark photon) field with a parity-odd coupling to the inflaton.  There are two regimes of production, delineated by the relative size of the coupling and the dark photon mass, arising from a tachyonic instability, and a parametric resonance, respectively. As a result of the coupling, the transverse modes, subdominant in both the minimally and nonminimally coupled theories, can dominate the total late-time abundance. The viable parameter space for dark-photon dark matter is thus widened. Furthermore, in the tachyonic case, the production is dominated by a single helicity mode, leading to a population of chiral dark photons that can lead to potential parity-violating observational signatures. We further explore the dependence of the production on the reheating history, and find that information about reheating is encoded in the number density spectrum.  

\end{abstract}

\date{\today}
\maketitle

\section{Introduction}
\label{sec:intro}
The universe is thought to have undergone an inflationary period of accelerated expansion in the early stages of its evolution, driven by a field known as the inflaton. One inevitable consequence of this rapid expansion of spacetime is cosmological gravitational particle production (CGPP), in which massive particles are created from the vacuum due to violation of adiabaticity during the expansion \cite{Parker1968, Parker:1969au, Ford:2021syk,Chung:1998zb,Chung:1998ua}. CGPP is a generic feature of quantum fields in curved spacetime and can produce particles of varying spins and over a wide mass range, as demonstrated in e.g., \cite{Chung:1998zb, Graham:2015rva, Ema:2015dka, Ema:2016hlw, Ema:2019yrd, Kolb:2020fwh,Ahmed:2020fhc, Alexander:2020gmv,Kolb:2021xfn,Kolb:2021nob,Kolb:2023dzp, Capanelli:2023uwv,Kaneta:2023uwi,Maleknejad:2022gyf,Capanelli:2024rlk,Capanelli:2024pzd,Jenks:2024fiu,Ling:2025nlw,Chowdhury:2025mye, Kolb:2026xfp}. Crucially, this production does not require any direct couplings to Standard Model (SM) particles or the inflaton, making CGPP a remarkably minimal process. As a result, it has been suggested as a production mechanism for dark matter, as well as a possible source for leptogenesis \cite{Bernal:2021kaj, Hashiba:2019mzm, Fujikura:2022udt, Flores:2024lzv, Chowdhury:2026zox}, and other cosmological observables. For further detail see the review \cite{Kolb:2023ydq} and references therein.

While the exact field content of the early universe is unknown, spin-1 vector fields, also known as dark photons, are a theoretically well motivated candidate to have been present. These fields can play a role in inflation, contribute to magnetogenesis, source primordial gravitational waves, and constitute some or all of the dark matter. The CGPP of massive vector fields has been studied in the minimally coupled case \cite{Graham:2015rva, Ema:2019yrd, Kolb:2020fwh, Ahmed:2020fhc} as well as the scenario in which a vector field has nonminimal couplings to gravity \cite{Alonso-Alvarez:2019ixv, Capanelli:2023uwv, Ozsoy:2023gnl, Cembranos:2023qph,Capanelli:2024pzd, Capanelli:2024rlk}. Less studied, however, is the CGPP of a massive vector field with a chiral coupling to the inflaton. Such an interaction is well studied in inflationary and preheating contexts, when the inflaton couples to a massless vector with the chiral Chern-Simons term e.g., \cite{Garretson:1992vt,Anber:2006xt,Anber:2009ua, Adshead:2015pva,Deskins:2013dwa}. Within this setup, it is known that the massless gauge fields are tachyonically produced and can impact late-time observables including primordial gravitational waves \cite{Dimastrogiovanni:2016fuu, Adshead:2018doq} and non-gaussianities \cite{Barnaby:2010vf, Agrawal:2018mrg, Dimastrogiovanni:2018xnn}. The CGPP of massive, chiral, dark photons shares many similarities with massless vector theory, and there is a smaller body of work on aspects of massive dark photons with this coupling to the inflaton \cite{Bastero-Gil:2021wsf, Ferreira:2026klg,Baker:2026xsd}.

In this work, we numerically compute the CGPP of dark photons $A_\mu$, with a chiral coupling to the inflaton $\varphi$ of the form $\varphi F \tilde{F}$. We find that the additional coupling to the inflaton amplifies the production of the transverse modes of the vector fields. Within this overall amplification, there are two regimes, depending on the strength of the coupling, which we refer to as the `Mathieu-like' regime and the tachyonic regime, for small and large couplings, respectively. In the Mathieu-like regime, the oscillations of the inflaton field imprint themselves as small oscillations in the dispersion relation of the vector field. This gives the equations of motion a similar form to a Mathieu equation, leading to a resonance that amplifies both helicity modes equivalently. On the other hand, when the chiral coupling is large, one of the helicity modes undergoes a tachyonic phase prior to the end of inflation, which leads to an overall chiral enhancement of one mode over the other. In both scenarios, the resultant number density of the transverse modes can become comparable to, or dominate over that of the longitudinal mode, which is the dominant contribution in both the minimally and nonminimally coupled scenarios. In both regimes, we explore the effects of reheating, and characterize the particle production in  early and late reheating scenarios. In the late reheating case, there is apparent `runaway' particle production at large momentum. When reheating occurs earlier, this runaway is controlled, and information about the time of reheating is encoded in the number density spectrum. As a result of the enhanced particle production, we find that viable parameter space for massive dark photon dark matter (DM) is widened compared to the minimal scenario; the correct dark matter abundance can be reached for a wide range of cosmologically reasonable parameters, for masses from $\mathcal{O}({\rm GeV})$ to the supermassive scale. Finally, we comment on the CGPP of massless fields in this theory, noting that due to the breaking of conformal invariance from the chiral term, CGPP is able to produce massless (or very light) fields, which can source primordial magnetic fields, in agreement with previous literature.

The structure of the paper is as follows. In Section~\ref{sec:chiralvec} we provide an overview of the theoretical setup of the massive vector theory and discuss the origins of the two resonance mechanisms. In Section~\ref{sec:GPPABC} we give a brief background on CGPP, then present our CGPP results for massive, chiral dark photons in the late and early reheating scenarios in Sections~\ref{sec:GPP-late}, and ~\ref{sec:Early}, respectively. We discuss the production of massless vectors in Section~\ref{sec:massless}, and the prospects for dark matter in Section~\ref{sec:DM}. Finally, we conclude with a discussion in Section~\ref{sec:conclude}. Throughout the paper, Greek indices refer to four spacetime dimensions, $i,j,k$ are spatial indices, and we use a mostly minus metric signature. 
\section{Chiral Vector Fields}
\label{sec:chiralvec}

We are interested in the gravitational production of massive, chiral dark photons (see \cite{Fabbrichesi:2020wbt} and references therein for a detailed dark photon review). The CGPP of dark photons has been studied in a variety of scenarios, including in the minimal theory \cite{Graham:2015rva,Ahmed:2020fhc,Kolb:2020fwh}, and with nonminimal couplings to gravity \cite{Alonso-Alvarez:2019ixv, Capanelli:2023uwv, Ozsoy:2023gnl, Cembranos:2023qph,Capanelli:2024pzd, Capanelli:2024rlk}. In the latter case, \cite{Capanelli:2024pzd,Capanelli:2024rlk} found that adding nonminimal couplings can introduce instabilities in the theory and lead to enhanced (or runaway) particle production. 

Nonminimal gravitational couplings are not the only natural extension of the minimal theory, however. Vector fields can also couple directly to pseudoscalars via a parity-odd Chern-Simons interaction. Such an interaction has been studied in inflationary settings, for example, \cite{Garretson:1992vt,Anber:2006xt,Anber:2009ua, Adshead:2015pva,Deskins:2013dwa}, where the pseudoscalar $\varphi$ is identified with the inflaton and couples to a massless vector field. It has also been studied in a late-time context, assuming the pseudoscalar is ultralight dark matter \cite{Aires:2026get, Kamali:2026tgq, Daniel:2026sik, Alexander:2026hgt}. It is known that in these scenarios, the chiral interaction leads to resonant production of the massless vector fields.  Motivated by these studies, we will consider a theory of a massive vector field, $A_\mu$, which has an axial coupling to the inflaton, $\varphi$, as follows:
\begin{align} 
S = & \int d^4 x \sqrt{-g}\left(-\frac{1}{4}F_{\mu\nu}F^{\mu\nu} + \frac{1}{2}m^2 A_\mu A^\mu 
\right. \nonumber \\
& \left. \hspace{60pt} + \frac{\alpha}{4f}\varphi F_{\mu\nu}\tilde{F}^{\mu\nu}\right),
\label{eq:action}
\end{align} 
where $\alpha$ is a dimensionless constant, $f$ is a constant with units $[M]$, and $\tilde{F}_{\mu\nu}$ is the Hodge dual of the field strength $F_{\mu\nu}$,
\be 
\tilde{F}^{\mu\nu} = \frac{1}{2}\epsilon^{\mu\nu\alpha\beta}F_{\alpha\beta}.
\ee 
We can decompose $A^\mu$ in terms of mode functions, $A^\mu_{\mathbf{k}}$ as 
\be 
A^\mu(t,\xvec) = \int \frac{d^3 \kvec}{(2\pi)^3}A^\mu_{\kvec }e^{i\kvec\cdot \xvec},
\ee 
where $\kvec$ is the wavenumber. The spatial components of $A^\mu_{\kvec}$ can then be decomposed into longitudinal and transverse polarization modes, $A_{\kvec}^{\rm L}$ and $A_{\kvec}^\pm$, where $\pm$ refers to the helicity of the two transverse components. With these expansions and assuming a cosmological FLRW background of the form
\be 
ds^2 = a^2(\eta)(d\eta^2 - d\mathbf{x}^2),
\ee 
where $\eta$ is conformal time and $a(\eta)$ the scale factor, the mode equations for the longitudinal and transverse modes, respectively, become
\begin{align}
    \partial_\eta^2 A_{\kvec}^L + \omega^2_L A_{\kvec}^L &=0,\\
    \partial_\eta^2 A_\kvec^{\pm} + \omega^2_\pm A_{\kvec}^{\pm} &= 0,
    \label{eq:modeqs}
\end{align}
with
\begin{align}
    \omega_L^2 &= k^2 + a^2 m^2 + \frac{1}{6}\frac{k^2}{k^2 + a^2m^2}a^2 R \nonumber \\ & \hspace{24pt} + 3 \frac{k^2}{(k^2 + a^2m^2)^2}a^4 H^2 m^2,\\
    \omega_\pm^2  &= k^2 + a^2 m^2 \pm \frac{\alpha}{f}\varphi^\prime k,
\end{align}
where $\prime$ denotes a derivative with respect to $\eta$, $H$ is the Hubble parameter, and $R$ the Ricci scalar.

Notice that the addition of the chiral coupling to the inflaton only impacts the transverse $A^\pm$ modes; the longitudinal modes remain unchanged. In the free theory, without any additional couplings, the CGPP of massive vectors is dominated by the longitudinal modes, and even when nonminimal couplings to gravity are added, only the longitudinal modes are affected. However, we will see that as a result of the chiral interaction, it is possible for the transverse modes to be the dominant contribution. 

It is useful to work in dimensionless units. In what follows, a parameter with a tilde will refer to the rescaling of the parameter to be dimensionless. For the inflationary background we take 
\begin{align}
\tilde{\varphi} = \frac{\varphi}{M_{\rm Pl}}, \quad & \frac{d}{dx} = \frac{d}{d(m_\varphi t)},
\end{align}
such that $\varphi' \equiv d\varphi/d\eta$ is given by 
\be 
\varphi' = m_\varphi M_{\rm Pl} a \frac{d\tilde{\varphi}}{dx} \,.
\ee 
The subscript `$e$' will be used to refer to the value of a parameter at the end of inflation.   The dimensionless conformal time is $\tilde \eta = a_eH_e\eta$.  Then, for the dimensionless $\omega$, defined as $\tilde \omega_k\equiv \omega_k/(a_eH_e) $, we have 
\be 
\tilde \omega_k ^2 = \frac{k^2}{(a_eH_e)^2} + \frac{a^2 m^2}{(a_eH_e)^2} \pm \frac{\alpha}{f}\frac{k}{(a_eH_e)^2}m_\varphi M_{\rm Pl} a \frac{d\tilde{\varphi}}{dx}\,.
\ee 
Finally, for numerical convenience, we define the dimensionless parameter $g$ such that 
\be 
g = \frac{{\alpha} m_\varphi M_{\rm Pl}}{ f H_e}.
\ee 
Then, the equations of motion that we numerically solve are: 
\be 
\partial^2_{\tilde\eta} A^{\pm}_{\kvec} + \left(\tilde{k}^2 + \tilde{a}^2\tilde{m}^2 \pm  32 g \tilde{a} \tilde{k}\frac{d\tilde{\varphi}}{dx}\right) A_{\kvec}^\pm = 0 \,, 
\ee
where $\tilde a = a/a_e, \tilde{m} = m/H_e$, and $\tilde{k} = k/(a_eH_e)$.

The theory in Eq.~\eqref{eq:action} is known to experience resonances in the gauge field sector as a result of the coupling to $\varphi$. There are two mechanisms that contribute to the amplification of $A_{\kvec}^\pm$. The first is a \textit{tachyonic resonance}, arising from the fact that $\varphi^\prime$ is fixed during the slow-roll phase prior to the end of inflation. Therefore, if $ g \tilde{k}\varphi^\prime > \tilde{k}^2 + \tilde{a}^2 \tilde{m}^2$, there will be a sustained period of negative $\omega_k^2$ for $A_{\kvec}^+$, corresponding to a temporary tachyonic phase causing the mode to get exponentially amplified. This mechanism is intrinsically chiral and leads to an overall asymmetry between the $A_{\kvec}^\pm$ modes. The tachyonic resonance regime was also studied in the context of CGPP in \cite{Bastero-Gil:2021wsf}.

The second mechanism is relevant when  $ g \tilde{k}\varphi^\prime < \tilde{k}^2 + \tilde{a}^2 \tilde{m}^2$. After inflation, the inflaton begins to oscillate about its minimum, leading to oscillations through zero in $\varphi^\prime$; as a result, the chiral term in $\omega_k^2$ also becomes periodic. This periodicity leads to a parametric resonance that affects both $A_{\kvec}^\pm$ equivalently, such that there is enhancement, but no chiral asymmetry. This mechanism can be understood in the context of a Mathieu equation, so we will refer to it as a \textit{Mathieu-like resonance}. In flat space, setting $a=1$, approximating $\varphi = \varphi_0 \sin(m_\varphi t)$, and letting $z = m_\varphi t/2$, allows us to rewrite Eq.~\eqref{eq:modeqs} as 
\be 
\partial_z^2 A^\pm_\kvec + \left[\bar{A}_k - 2 q \cos (2z)\right]A^\pm_\kvec = 0, 
\label{eq:flat-mathieu}
\ee 
where 
\begin{align}
\bar{A}_k = \frac{4(k^2 + m^2)}{m_\varphi^2} \, ,\quad q = \mp 2 \frac{\alpha}{f}\frac{\varphi_0 k}{m_\varphi}.
\end{align}

This is exactly the Mathieu equation, which has solutions of the form $A^\pm_\kvec \propto e^{\mu_k z}$, where $\mu_k$ are the Floquet exponents, that depend on $q$ and $\bar{A}_k$. The solutions trace out bands in the parameter space in which a given mode will undergo parametric resonance when when $\mu_k >0$. 

While the flat space analysis provides intuition for the origins of the resonance, it does not completely capture the effects in an expanding spacetime. In an expanding background, both $\bar{A}_k$ and $q$ are time evolving. The former grows with $\bar{A}_k\propto a^2 m^2$, and the amplitude of $q$ decreases from the redshifting of $\varphi^\prime$, and the mode equation solutions are no longer a clean exponential. As a result, each mode no longer sits at a point within or outside of a resonance band in the parameter space, but rather traces out a trajectory through multiple bands as it evolves. This was first discussed in the context of reheating in \cite{Traschen:1990sw, Kofman:1994rk, Kofman:1997yn, Greene:1997fu}. Ref. \cite{Kofman:1997yn} characterized the mode evolution in an expanding spacetime as a broad, stochastic parametric resonance, during which modes are amplified beginning at large $q$ as they pass through the bands (and adiabaticity is violated), and evolve adiabatically in between. The amplification continues and as the expansion of the universe slows down, the modes spend longer in each band, more closely approximating the flat-space narrow resonance, until the mode exits the resonance regime all together.

\section{Gravitational Particle Production}
\label{sec:GPPABC}

In this section we provide a brief overview of the mechanics of CGPP. An old idea, originated by Schr\"{o}dinger \cite{Schrodinger1939}, and formulated in the context of modern field theory by Parker \cite{Parker1968}, CGPP is based on the simple idea that the rapid, non-adiabatic expansion of spacetime during inflation creates particles from the vacuum. It was suggested as a production mechanism for dark matter by Chung, Kolb, and Riotto \cite{Chung:1998zb}, and can have other cosmological consequences during inflation and at late times (for a review, see \cite{Kolb:2023ydq}).

For representative purposes, we choose to study CGPP in a quadratic inflation model, assuming that the inflaton dynamics are governed by:
\be 
\ddot{\varphi} + 3 H \dot{\varphi} + m_\varphi^2\varphi = 0.
\ee 
The inflaton slowly rolls prior to the end of inflation, then begins to oscillate at the bottom of its potential. This leads to a post-inflationary matter-dominated era, followed by reheating as the oscillations cease and the inflaton decays into SM particles.

Well before the end of inflation, at early times, we can define an initial vacuum state with early-time creation and annihilation operators, $\hat{a}_\kvec^\dagger$ and $\hat{a}_\kvec$, as 
\be 
\hat{a}_\kvec^\early  \ket{0^\early} = 0\ket{0^\early }. 
\ee 
These operators are related to the post-inflationary, late-time operators as 
\begin{align}
\hat{a}_\kvec^\early = \alpha_\kvec^* \hat{a}_\kvec^\late - \beta_\kvec^* \hat{a}_{-\kvec}^{\late\dagger},
\end{align}
where $\alpha_k$ and $\beta_k$ are known as the Bogoliubov coefficients. Upon solving the mode equations, Eq.~\eqref{eq:modeqs}, $\beta_k$ is constructed at late times as follows: 
\be 
\label{eq:betak}
|\beta_{k}|^2 = \lim_{\eta \rightarrow \infty} \left[\frac{\omega_k^\lambda}{2}|A^\lambda_\kvec|^2 + \frac{1}{2\omega_k^\lambda}|\partial_\eta A^\lambda_\kvec|^2 - \frac{1}{2}\right], 
\ee 
where $\omega_k$ is the frequency of the mode and $\lambda  = $\{L, + , -$ \}$ denotes the mode. 

The number density spectrum in terms of the comoving wavenumber, $k$, can be constructed in terms of $\beta_k$  as \cite{Kolb:2020fwh}
\begin{equation}
n_k =\frac{k^3}{2\pi^2}|\beta_k|^2. 
\label{eq:nk}
\end{equation}
Finally, the comoving number density is given by integrating over Eq.~\eqref{eq:nk}:
\be 
na^3 = \int \frac{dk}{k} n_k.
\label{eq:na3}
\ee 

In what follows, we will follow this procedure to compute the particle production.

\section{Production in Late Reheating}
\label{sec:GPP-late}
Having discussed the relevant background details, we now turn to our main results. In this section we discuss CGPP of massive dark photons in the late reheating scenario, and characterize the two resonance regimes. We focus primarily on the production of the transverse modes of the vector field, given that the longitudinal modes are unchanged by the coupling to the inflaton. For late reheating, we assume that reheating occurs well after the end of the particle production era, such that the quasi-de Sitter inflationary epoch transitions into a matter dominated universe. We will discuss the alternative, early reheating, in Section~\ref{sec:Early}. For notational simplicity, we will now drop the `$\kvec$' subscript in the mode functions and refer to the transverse modes as $A_\pm$ going forward.

\begin{figure}[htb!]
    \centering
    \includegraphics[width=0.48\textwidth]{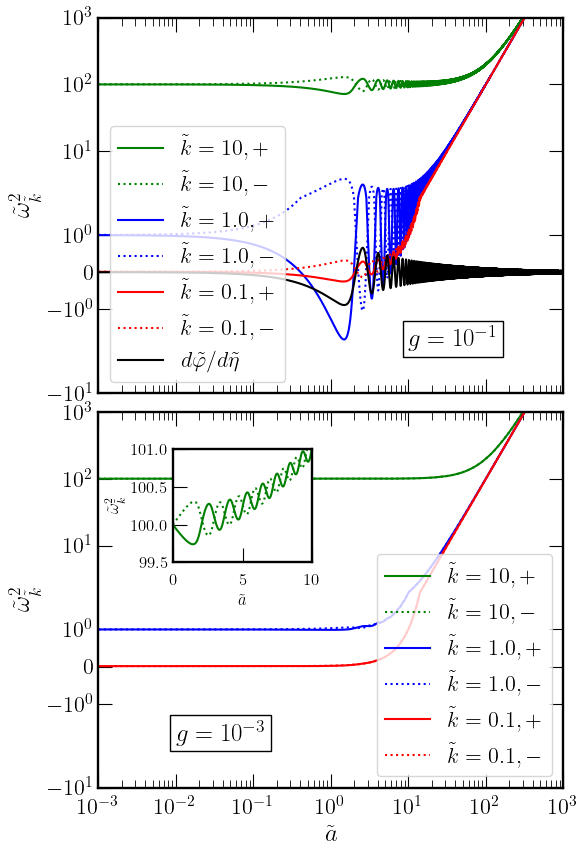}
    \caption{The dispersion relation, $\tilde\omega^2_{\tilde k}$, for $\tilde m=0.1$ and $g=10^{-1}$ (top figure) and $g = 10^{-3}$ (bottom figure), comparing the plus and minus modes for various $\tilde{k} = 0.1, 1$, and $10$.}
    \label{fig:omegaksq_1}
\end{figure}

Independent of when reheating occurs, there are two distinct regimes of particle production, arising from the two resonance mechanisms discussed earlier in Section~\ref{sec:chiralvec}. Which mechanism dominates the particle production depends on the ratio of the coupling $g$ to the mass $\tilde{m}$. For large $g/\tilde m$, particle production is amplified by the tachyonic resonance in one of the $\pm$ modes, while for small $g/\tilde m$, the tachyonic resonance is less relevant, but the production of both modes is still enhanced from the Mathieu-like resonance. This distinction stems from the behavior of $\tilde{\omega}_k^2$ in these two regimes. Figure~\ref{fig:omegaksq_1} shows $\tilde{\omega}_k^2$ for fixed $\tilde{m} = 0.1$; we show small ($\tilde{k} = 0.1$), intermediate ($\tilde{k} = 1$), and large ($\tilde{k} = 10$) $\tilde{k}$ values for both transverse helicity modes. The top panel shows $g = 0.1$, where at small and intermediate $\tilde{k}$ the $A_+$ mode has an extended tachyonic phase prior to the end of inflation, then oscillates with $\tilde{\varphi}^\prime$ at late times, while the $A_-$ mode remains non-tachyonic until inflation ends and it begins oscillating through $\tilde{\omega}_k^2 = 0$. At high \kt, neither mode becomes tachyonic, though both undergo small oscillations due to $\tilde{\varphi}^\prime$. This behavior leads to an amplification of both the $A_\pm$ modes compared to the minimal theory, with the $A_+$ mode further enhanced by its tachyonic phase during inflation.

In the lower panel, we show the same three modes for $\tilde{m} = 0.1$ and $g = 10^{-3}$. Here, there is no appreciable tachyonic contribution, as $\tilde{\omega}_k^2$ is dominated by the non-chiral terms. However, small oscillations persist, as can be observed in the inset. Despite the small amplitude of these oscillations, there will still be an overall enhancement that affects both $A_\pm$ equivalently, arising from the Mathieu-like resonance. 

\begin{figure}[htb!]
    \centering
     \includegraphics[width=0.48\textwidth]{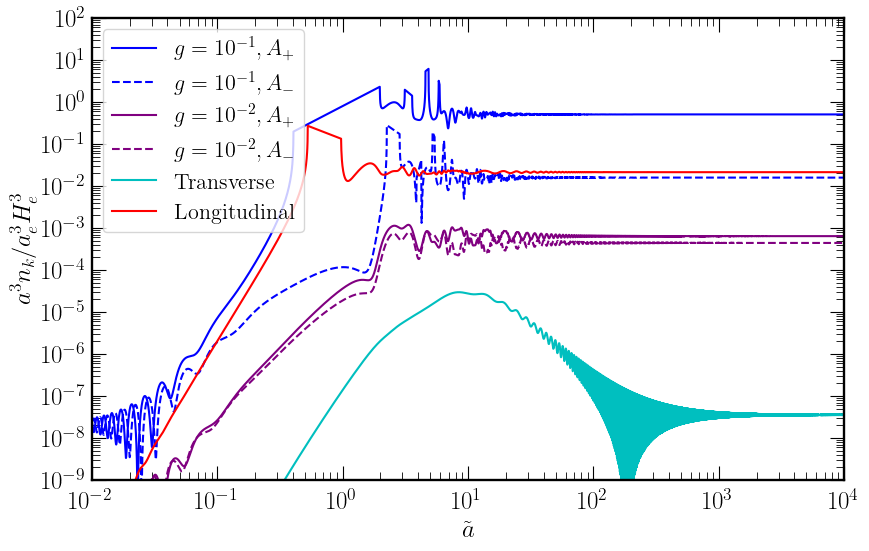}
   \caption{Comparing the time evolution of $a^3 n_k/(a_e H_e)^3$ for $A_\pm$ to $A_{\rm L}$ and the $g=0$ transverse mode. We fix $\tilde m=0.1$ and $\tilde k = 1$ and choose $g=10^{-1}, \ 10^{-2}$ as representative examples.}
    \label{fig:mode-evol}
\end{figure}

To explicitly observe the effects in these two regimes, consider the evolution of the number density of several different modes. Figure~\ref{fig:mode-evol} shows the evolution of a mode with $\tilde m = 0.1, \tilde{k}=1$ as a representative example. We compare the evolution of the $g=0$ minimal transverse $A_{\rm T}$ (cyan) with $A_\pm$ (solid and dashed, respectively) for $g=0.1$ (blue) and $g=10^{-2}$ (purple). We also show the longitudinal mode, $A_{\rm L}$ (red), which is unaffected by the chiral coupling, but is the dominant contribution to the number density in the non-chiral theory.

We can appreciate the distinction in the evolution of each of the modes. For $g=0.1$ in the tachyonic regime, there is a substantial enhancement in the number density compared to that of the $g=0$ transverse mode, $A_{\rm T}$, and an enhancement of nearly two orders of magnitude of $A_+$ over $A_-$. Furthermore, both of the chiral transverse modes are dominant over the longitudinal mode. On the other hand, for $g= 10^{-2}$, which is much closer to the Mathieu-like resonance regime, $A_+$ and $A_-$ are nearly equally enhanced over $A_{\rm T}$.

\begin{figure}[htb!]
    \centering 
\includegraphics[width=0.48\textwidth]{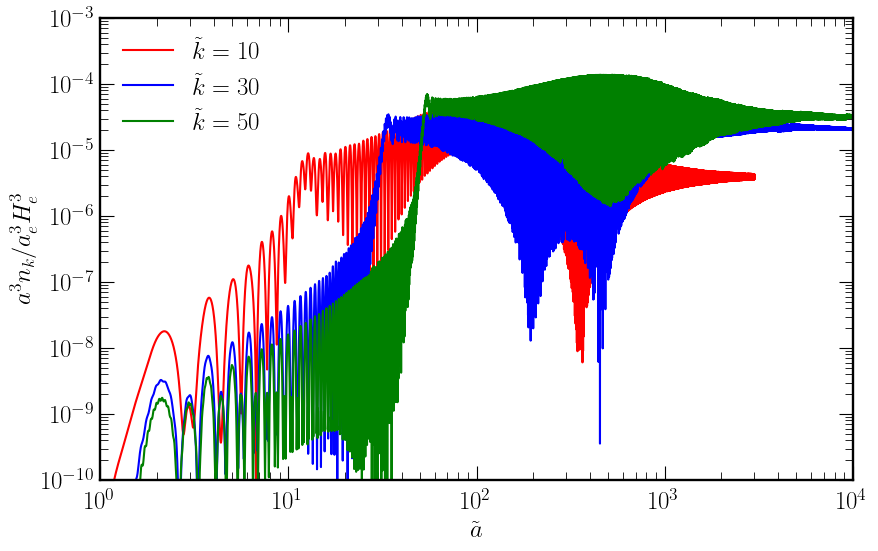}
\caption{High-$k$ mode evolution for $A_+$, $\tilde m=0.1, g=10^{-4}$ in the Mathieu-like regime.
}
\label{fig:nk-a-mup1g1em4}
\end{figure}

Let us consider the behavior of the Mathieu-like resonance in further detail. Given that the chiral term is proportional to $\tilde k$, it is instructive to study the evolution of the high-\kt\ modes. Figure~\ref{fig:nk-a-mup1g1em4} shows the evolution of $a^3 n_k/a_e^3H_e^3$ of $A_+$ for $\tilde{k} = 10,\ 30$, and $50$ with $\tilde m = 0.1$ and $g = 10^{-4}$, well within the Mathieu-like regime. For these parameter values, the amplitude of oscillations is  small, for example at $\tilde k = 10$, the maximum oscillation peak is less than half a percent from the minimal baseline. Despite the small amplitude, however, the oscillations clearly impact the number density in a dramatic way. 
Particularly for $\tilde k = 30, 50$, the sharp, drastic rise in $a^3n_k/a_e^3H_e^3$ onsets well after the end of inflation and terminates shortly thereafter, indicating that the mode is in the narrow regime of the Mathieu-like resonance.  

\begin{figure}[htb!]
    \centering
\includegraphics[width=0.48\textwidth]{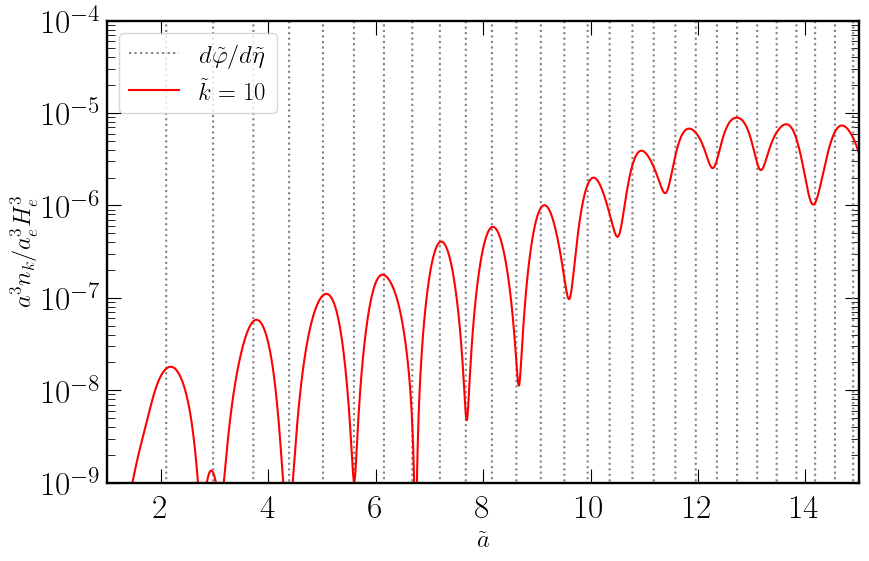}
\includegraphics[width=0.48\textwidth]{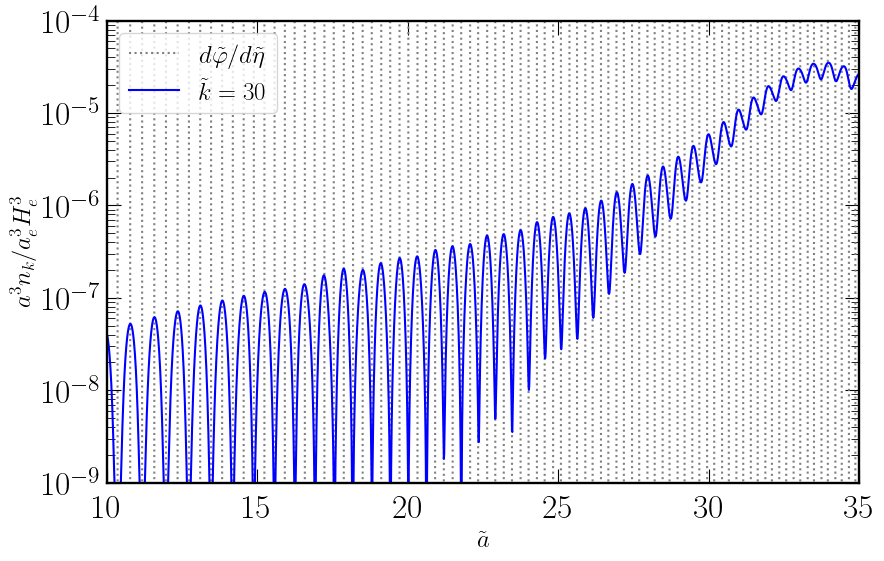}
\includegraphics[width=0.48\textwidth]{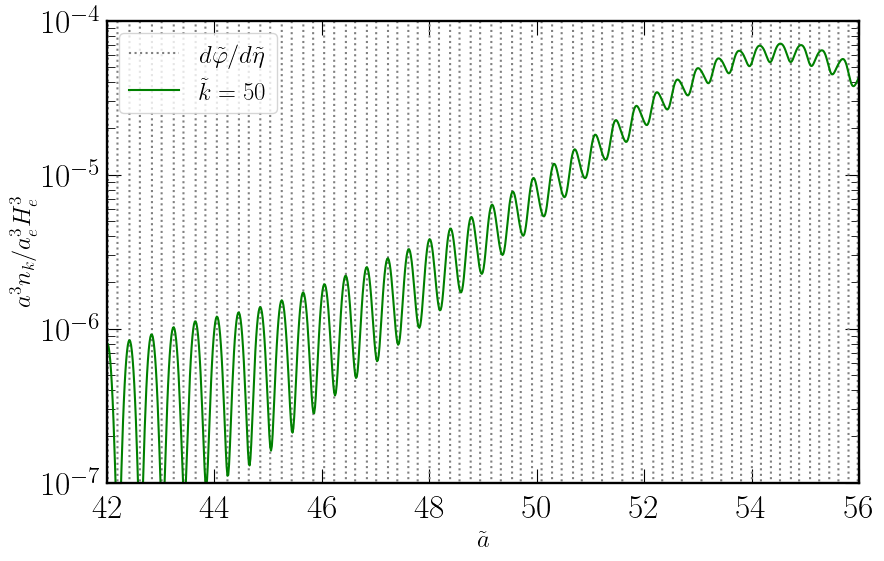}
\caption{Number density evolution for $A_+$, $\tilde m=0.1, g=10^{-4}$, with $d\tilde \varphi/d\tilde \eta$ overlaid with $a^3n_k/a_e^3H_e^3$. We zoom in at late times near the sharp increases in number density for $k=10,\ 30,\ 50$. We can see a clear alignment of the frequency of $a^3n_k/a_e^3H_e^3$ with that of $d\tilde \varphi/d\tilde \eta$ until $\tilde a \sim \tilde k$, at which point the frequencies dephase.   }
\label{fig:resonancezoom}
\end{figure}

To further emphasize this, in Fig.~\ref{fig:resonancezoom}, we show a zoom-in of $a^3 n_k/a_e^3H_e^3$ for each of the three modes superimposed with $\varphi^\prime$ in the resonance regime. For each of the modes, the frequency of $a^3 n_k/a_e^3H_e^3$ has near perfect alignment with the background $\varphi^\prime$, until $\tilde a \sim \tilde k$, at which point the frequencies dephase, the mode exits the resonance regime and eventually stabilizes to its final asymptotic value. This is in contrast to the minimal theory, in which the number densities of the high-\kt \, modes are suppressed beyond $\tilde k \sim \tilde m$.

\begin{figure}[htb!]
    \centering
\includegraphics[width=0.48\textwidth]{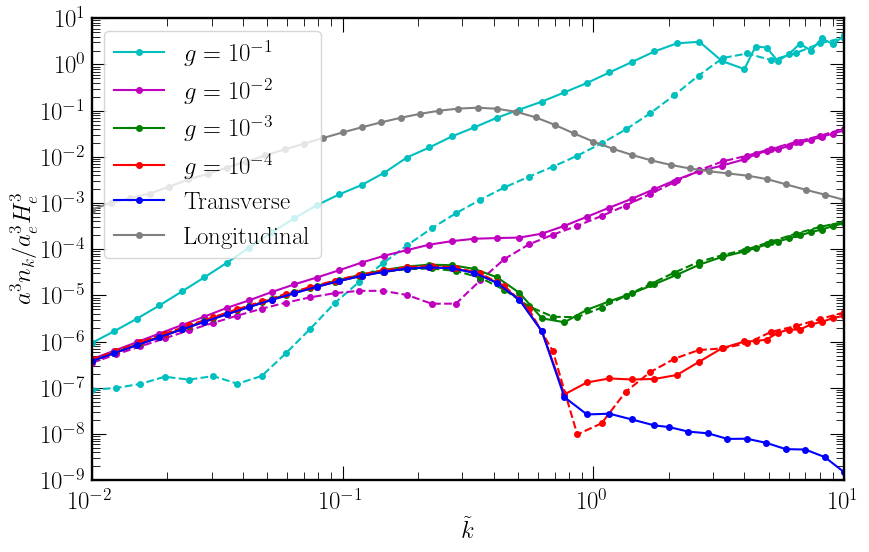}
\caption{Number density spectra for $A_\pm$, with $\tilde m=0.1$ and varying $10^{-4} < g < 10^{-1}$. The solid (dashed) curves are for $A_+$ ($A_-$), while the blue and grey curves show the $g=0$ transverse and longitudinal spectra, respectively. }
\label{fig:nk-spectrum-varyg}
\end{figure}

To illustrate full impact of the chiral coupling and its effect on the high-\kt \, modes, we now turn to the number density spectrum of $A_\pm$ in terms of \kt, taking the asymptotic, late-time value from the time evolution of each mode. Figure~\ref{fig:nk-spectrum-varyg} shows $a^3 n_k/a_e^3H_e^3$ for \mt = 0.1 and $10^{-4}\leq g \leq 10^{-1}$. For the smallest couplings, $g=10^{-3}$ and $10^{-4}$, the number densities of $A_\pm$ tracks the $g=0$ transverse spectrum until $\tilde{k} \sim 1 $, beyond which they begins to grow due to the Mathieu-like resonance from the small oscillations in $\tilde{\omega}_k^2$. For $g = 10^{-2}$, the spectra begins diverging from $g=0$ at lower \kt \, and grow more drastically, while for $g=0.1$, the growth in the number density is the most pronounced, exceeding the $g=0$ number density by nearly ten orders of magnitude at $\tilde{k} =10$. In both of these cases, the $A_-$ number density behaves similarly, but with a lower overall amplitude than that of the $A_+$ mode early in the evolution, when the number density is dominated by the tachyonic evolution. 

A concern with the results in Figures~\ref{fig:nk-a-mup1g1em4} and~\ref{fig:nk-spectrum-varyg} is that $a^3n_k/(a_e H_e)^3$ appears to diverge at large $\tilde{k}$ due to the continued oscillations of the inflaton. For a physical interpretation of the CGPP, there must be some mechanism to tame this growth. Fortunately, the growth is not a true runaway, given that the production ceases when the universe reheats. We will show this explicitly in the next section when discussing early reheating. It is also possible that once the number density becomes large enough that backreaction of the gauge field production onto the inflaton dynamics becomes relevant, the seeming unbounded growth in $a^3n_k/a_e^3 H_e^3$ will cease. 

\section{Production in Early Reheating}
\label{sec:Early}

In the above discussion, we assumed a late reheating scenario, such that the inflaton continues to oscillate well after the end of inflation during a matter-dominated era. As a result, the Mathieu-like resonance continues even for large-\kt\, modes, which would otherwise have suppressed production, leading to a problematic runaway effect. To resolve this and explore the effects of an earlier exit out of the inflationary era, we now consider an `early reheating' scenario,\footnote{Early reheating is delineated by $H_\mathrm{RH}> m$, where $H_\mathrm{RH}$ is the expansion rate at the transition to the radiation-dominated era.  This criterion is equivalent to $\tilde a_\mathrm{RH} <\tilde m^{-2/3}$} in which the particle production terminates earlier. As a representative example, we take $\tilde a_{\rm RH} = 3.34$ and $\tilde m =0.1$.

\begin{figure}[htb!]
    \centering
    \includegraphics[width=0.95\linewidth]{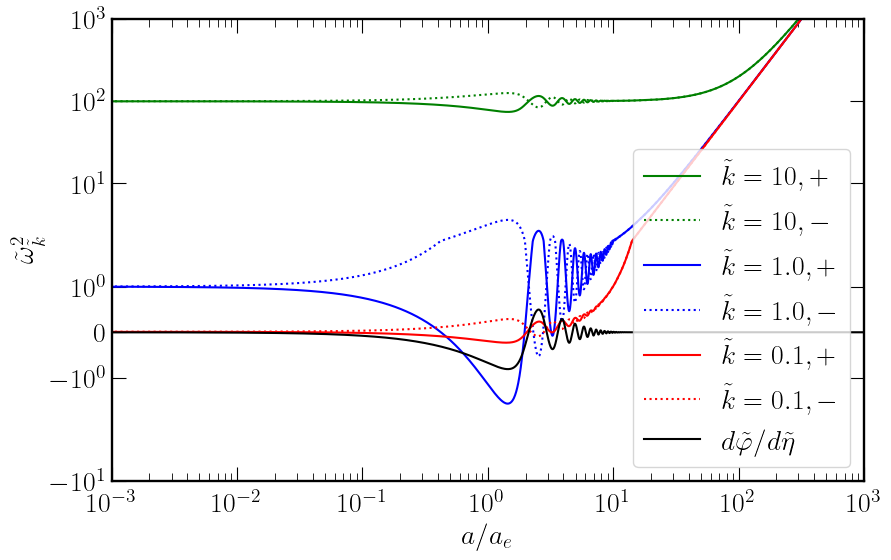}
    \caption{$\tilde \omega_{\tilde k}^2$ for early reheating, taking $\tilde m=0.1$ and $g=0.1$.
    }   
    \label{fig:omega-earlyRH}
\end{figure}

First, consider again the evolution of $\tilde{\omega}_k^2$. Figure~\ref{fig:omega-earlyRH} shows $\tilde \omega_{\tilde k}^2$ in the early reheating scenario for the same three $\tilde{k} = 0.1, 1, 10$ modes as in Figure~\ref{fig:omegaksq_1}. The behavior is similar to that in Figure~\ref{fig:omegaksq_1}: the low and intermediate $\tilde{k}$ $A_+$ modes undergo a tachyonic phase, and all experience post-inflationary oscillations. However, the amplitude of oscillation is smaller, and the duration is shorter relative to late reheating. This is because the inflaton decays into SM radiation much earlier than in the late reheating scenario, terminating the field oscillations earlier, and in turn shortening the resonant particle production regime. 

\begin{figure}[htb!]
\centering 
    \includegraphics[width=0.48\textwidth]{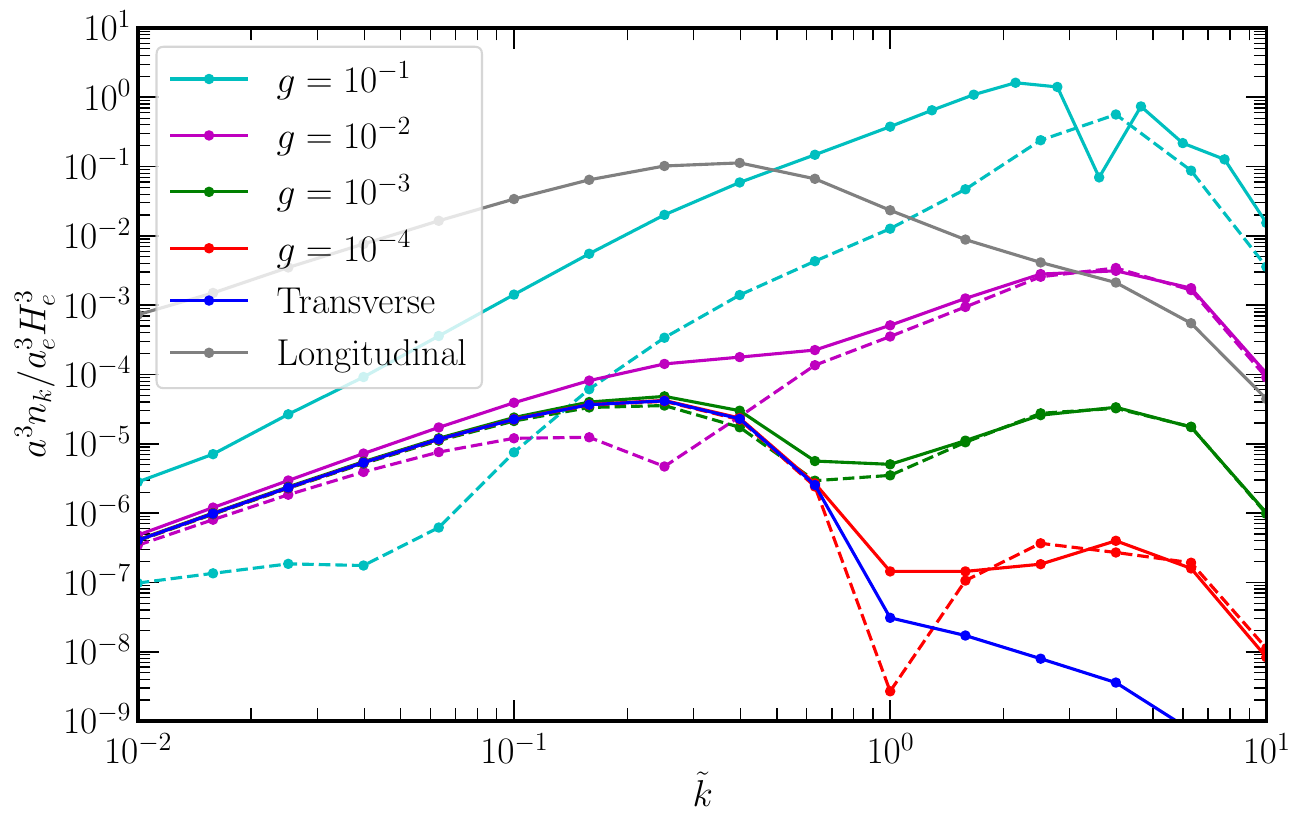}
    \caption{Particle number density spectrum for early reheating, $\tilde m =0.1$, $\tilde{a}_\mathrm{RH}=3.34$. We show the $A_+$ mode (solid) and the $A_-$ mode (dashed) for $10^{-4} < g < 10^{-1}$.}
    \label{fig:nkEarly}
\end{figure}

The consequences are apparent in the spectrum, shown in Figure~\ref{fig:nkEarly} for the $A_\pm$ number densities. There is still an enhancement relative to the $g=0$ modes, but approximately an order of magnitude smaller than in the late reheating scenario. As before, for small values of $g$ the particle production is dominated by the Mathieu-like resonance, amplifying both $A_\pm$ equally. For large $g$, the enhancement is dominated by the early-time tachyonic behavior of the field and leads to a preferential production of one chiral mode. A crucial difference between the early and late reheating scenarios is that the early reheating spectrum has a defined peak, and the growth in $n_k$ with $\tilde{k}$ seen in late reheating is controlled. Notably, for large values of $g$, the primary peak of the spectrum is at $\tilde k \sim  \tilde{a}_{\rm RH}$, indicating that the spectrum encodes information about the reheating epoch. Even for small values of $g$, the $\tilde k \sim \tilde{a}_{\rm RH}$ peak is secondary to that at $\tilde k \sim \tilde m$, but the structure of the spectra are still distinct from that of the $g=0$ transverse modes.

\begin{figure}[htb!]
   \centering
   \includegraphics[width=0.48\textwidth]{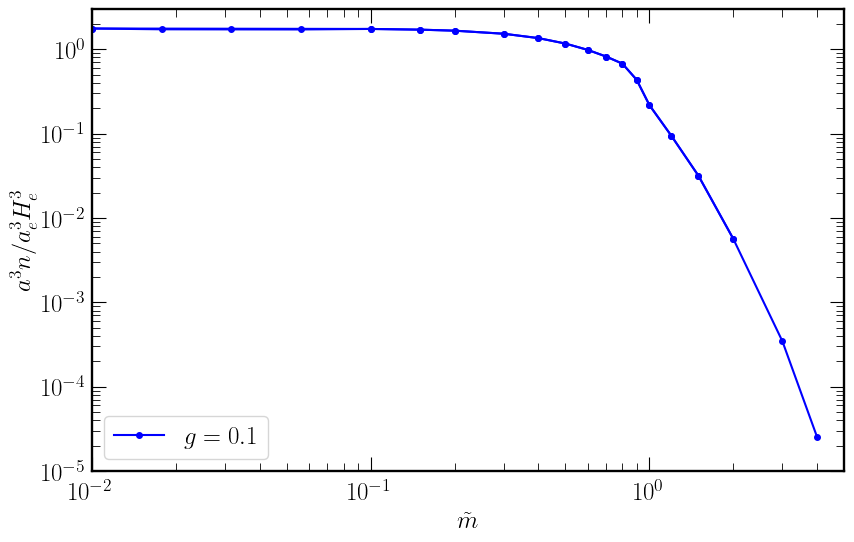}
    \caption{Total comoving number density $a^3n/a_e^3H_e^3$ as a function of $\tilde m$ for $A_+$ assuming early reheating, and $g=10^{-1}$.
   }
    \label{fig:na3-early}
\end{figure}

From the spectra, we can compute the comoving particle number density from Eq.~\eqref{eq:na3}, integrating from $10^{-2} \leq {\tilde k} \leq 10$. In Figure~\ref{fig:na3-early}, we show $a^3 n/a_e^3H_e^3$ in terms of the particle mass, $\tilde{m}$, for the $A_+$ mode with $g=0.1$. For this choice of $g$, the total number density is always dominated by the tachyonic production and therefore can be well approximated by the abundance of $A_+$ alone.  
At low mass, production is dominated by the chiral term in $\omega_k^2$. As a result, for $0 \lesssim \tilde{m}\lesssim 10^{-1}$, the comoving number density has negligible dependence on the mass and remains approximately constant. This is in contrast to the non-chiral theory, in which the number density is suppressed for small mass. As $\tilde{m}$ increases, the effect of the chiral term becomes less pronounced and $a^3n/a_e^3H_e^3$ begins to exponentially drop, with $a^3 n_k/a_e^3H_e^3 \propto e^{-2.8 \tilde{m}}$, in agreement with the known exponential suppression in the production of particles with $\tilde m > 1$; see e.g.,  \cite{Jenks:2024fiu}.

In the above discussion, we have chosen a particular value of $\tilde{a}_{\rm RH} = 3.34$ as a representative example, but the observed features in the CGPP remain for other choices of the reheating epoch as well. If $\tilde{a}_{\rm RH} > 3.34$, the inflaton oscillations continue longer after the end of inflation, leading to continued enhancement at high-$\tilde{k}$. As $\tilde{a}_{\rm RH} \rightarrow \infty$, the spectrum will smoothly approach the runaway observed in late reheating. On the other hand, if $\tilde{a}_{\rm RH} < 3.34$, the secondary peak will move to smaller $\tilde k$, and the production from the Mathieu-like resonance will eventually shut off, as instantaneous reheating is approached, leaving only the tachyonic enhancement.

\section{Production of Massless Fields}
\label{sec:massless}
In the previous sections, we discussed the production of massive vector fields. However, the explicit breaking of conformal invariance, due to the coupling of $A_\mu$ to the inflaton, also allows for the production of massless vector fields. This is distinct from the minimal CGPP scenario, which preferentially produces particles with masses near the Hubble scale. The production of massless fields from the parity-odd vector-inflaton coupling is well known in the context of inflation and preheating, as discussed in Sections~\ref{sec:intro} and~\ref{sec:chiralvec}. Here for completeness, we formulate it in the language of CGPP and connect to our massive dark photon results. 

\begin{figure}[htb!]
    \centering
    \includegraphics[width=0.48\textwidth]{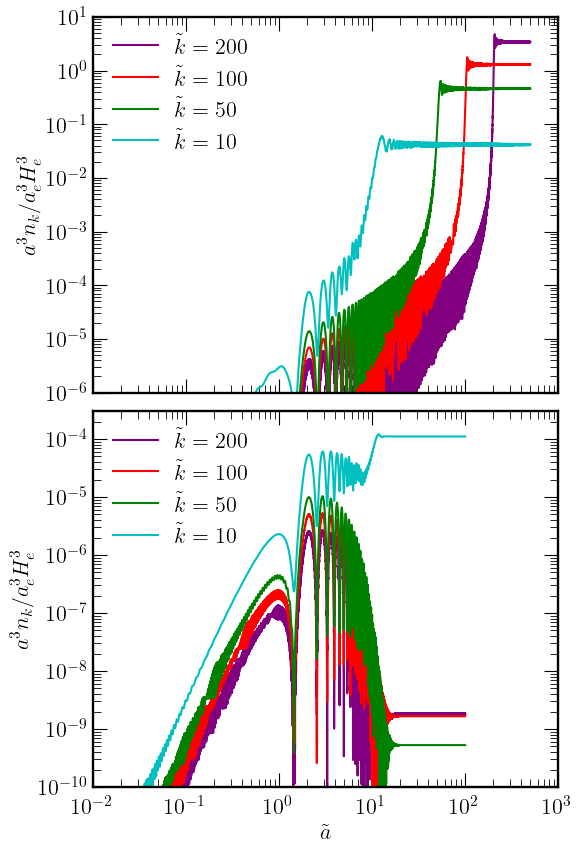}    \caption{Mode evolution for the $A_+$ mode with $\tilde m = 0$ and $g=10^{-2}$, for late (upper panel) and early (lower panel) reheating.
    }
    \label{fig:massless}
\end{figure}

For massless fields, the longitudinal mode vanishes and we are left with only the two transverse modes, which obey Eq.~\eqref{eq:modeqs} with $m =0$. From Eq.~\ref{fig:na3-early}, we expect that the particle number density will be maximized for $\tilde{m}=0$, given that the vanishing mass maximizes the relative contribution of the chiral term to $\tilde{\omega}_k^2$. This can be seen by comparing the high-$\tilde{k}$ modes in both the early and late reheating scenarios. Figure~\ref{fig:massless} shows this comparison for $\tilde{k} = 10,\, 50, \,100,\, 200$ for late reheating (top) and early reheating (bottom). The effects of the coupling are particularly visible at these large values of \kt. In the late reheating scenario, massless fields continue to be produced for increasing values of $\tilde{k}$. The oscillations of the inflaton lead to a sharp increase in the number density after the end of inflation up to $\tilde{a} \sim \tilde{k}$. On the other hand, in early reheating, the earlier cessation of the inflaton oscillations lead to a suppression in the number density at high $\tilde{k}$. However, there is still an enhancement compared to the $g=0$ scenario, which can be appreciated from the larger value of the number density for $\tilde{k}=100$ and $\tilde{k} = 200$ compared to $\tilde{k} = 50$. This is because there is a tradeoff between the usual high-$\tilde{k}$ suppression of the number density and the enhancement of the chiral term due to increasing $\tilde{k}$. Nevertheless, even with this enhancement, the total number density at large \kt\, remains well below the peak and there is no concern about a runaway effect.  

\begin{figure}[htb!]
    \includegraphics[width=0.48\textwidth]{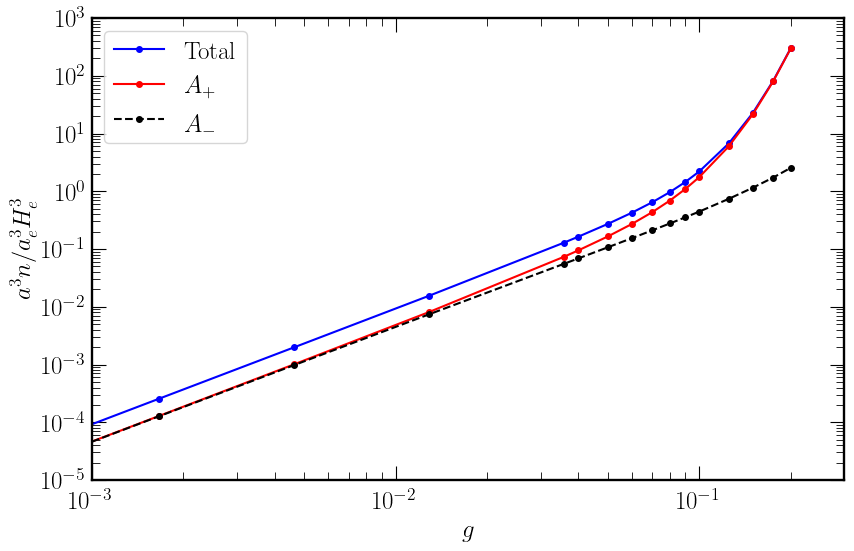}
    \caption{$a^3n/a_e^3H_e^3$ as a function of $g$ for $\tilde m=0$. We show the number density for $A_\pm$ (red, black dashed) as well as the total (blue). }
    \label{fig:varyg}
\end{figure}

Given that the comoving particle number is maximized for $\tilde{m}=0$, it is also useful to consider how the total number density varies with the strength of the coupling $g$, which sets an upper limit to the number density of \textit{massive} particles that can be produced for a given $g$. In Figure~\ref{fig:varyg} we show the total maximum comoving number density for the $A_+$ (red) and $A_-$ (black dashed) modes as well as the total (blue), assuming early reheating. Here we can explicitly see the effects of the Mathieu-like resonance amplifying both modes for $g \lesssim 10^{-2}$, where the number density scales as $g^2$. Then the production becoming dominated by the tachyonic resonance in the $A_+$ mode for large $g$, where the production begins exponentially rising. We do not consider backreaction of $A_\pm$ onto $\varphi$ and therefore only extend our analysis to modest values of $g$. For a full discussion of axion inflation including backreaction see e.g., \cite{Adshead:2015pva}. 

One cosmological application of the CGPP of massless vectors is the possibility of producing cosmologically relevant primordial magnetic fields. Our universe is replete with magnetic fields on a variety of scales, but the origin of galactic and intergalactic magnetic fields remains a mystery. Magnetogenesis in the early universe during the inflationary epoch is an attractive option and there are many models that have proposed generating magnetic fields by the axial coupling of a massless gauge field to the inflaton, e.g., \cite{Garretson:1992vt, Anber:2006xt,Byrnes:2011aa, Caprini:2014mja, Adshead:2016iae,Gorbar:2021zlr}. In these models, magnetic fields resonantly produced due to the chiral coupling to the inflaton can plausibly account for the observed present-day magnetic fields.

\section{Cosmological Implications for Dark Matter}
\label{sec:DM}

In the previous sections, we characterized the number density of chiral vectors produced during the inflationary epoch. One crucial question is whether these particles can be produced in the appropriate abundance to account for some or all of the dark matter in the present-day universe. Dark photons are a compelling dark matter candidate \cite{Fabbrichesi:2020wbt}. The possibility of CGPP as a dark matter production mechanism has been explored in a variety of scenarios \cite{Graham:2015rva, Ema:2019yrd, Kolb:2020fwh, Ahmed:2020fhc,Alonso-Alvarez:2019ixv,Bastero-Gil:2021wsf, Capanelli:2023uwv, Ozsoy:2023gnl, Cembranos:2023qph, Capanelli:2024pzd, Capanelli:2024rlk}. When the dark photon is beyond minimally coupled, for example with non-minimal couplings to gravity, there is an enhancement in production and therefore a widening of the viable DM parameter space. In our scenario, we expect that the additional chiral coupling to the inflaton will similarly open the parameter space.

From the comoving number density, $a^3n$, one can determine the relic abundance of dark matter today with \cite{Kolb:2023ydq}
\be 
\frac{\Omega h^2}{0.12} = \frac{m}{H_e}\left(\frac{H_e}{10^{12} {\rm GeV}}\right)^2\left(\frac{T_{\rm RH}}{10^{9} {\rm GeV}}\right)\frac{a^3n/a_e^3H_e^3}{10^{-5}},
\ee 
where $\Omega h^2/0.12 \sim 1$ corresponds to the observed DM abundance in our present-day universe, and $T_{\rm RH}$ is the reheating temperature. 

There are several distinctions between the minimal dark photon dark matter scenario and ours, in which the dark photon is produced with a chiral coupling to the inflaton. In the minimal theory with no additional couplings, the longitudinal mode is the dominant contribution to the particle number density for dark photon DM. However, with the additional chiral coupling to $\varphi$, the number density can primarily come from the transverse modes. This can be appreciated from Figure~\ref{fig:nk-spectrum-varyg}, where the total number density in the transverse modes becomes comparable to that in the longitudinal modes around $g \sim 10^{-2}$ for $\tilde{m} = 0.1$. In particular, for larger values of $g$ for which the number density is dominated by the tachyonic enhancement, there will also be a preferential chiral production of the $A_+$ mode. As a result, the dark matter can have an overall chirality and can impact cosmological observables. For example, the formation of polarized vector solitons and the collapse of chiral vector dark matter into halos has been explored in~\cite{Jain:2021pnk} and~\cite{Alexander:2026fam}, respectively. 

\begin{figure}[htb!]
    \includegraphics[width=0.48\textwidth]{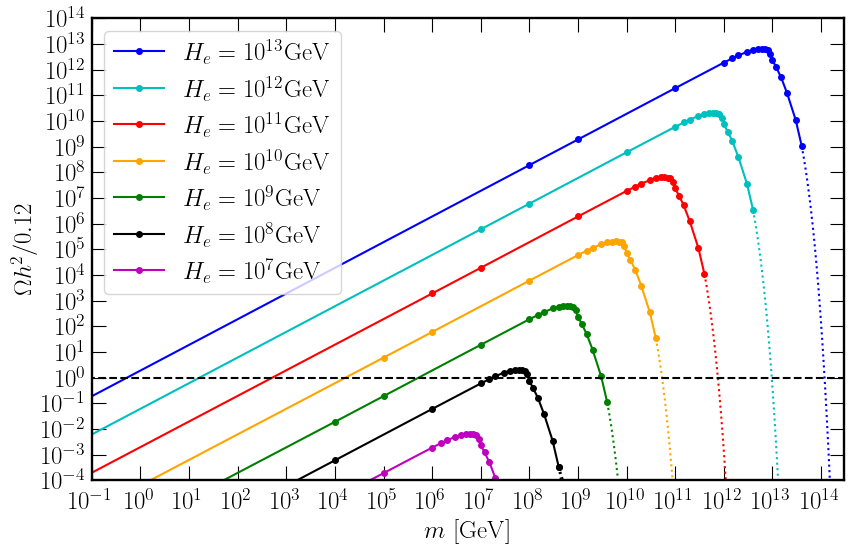}
    \caption{The dark matter relic density $\Omega h^2/0.12$ as a function of $m$ for $10^7 < H_e/{\rm GeV} < 10^{13}$ for $g=0.1$ in the early reheating scenario. The points are the result of numerical calculation and the dotted curves are extrapolation of the numerical results following $a^3n/a_e^3H_e^3 \propto e^{-2.8\tilde m}$.
    }
    \label{fig:DM}
\end{figure}

More broadly, the overall enhanced CGPP widens the parameter space in which CGPP can successfully produce massive dark photon dark matter compared to the minimal theory. Figure~\ref{fig:DM} shows $\Omega h^2/0.12$ as a function of the physical DM mass, $m$, taking $g=0.1$ and the early reheating scenario as a representative example of the parameter space. For early reheating, $T_{\rm RH}$ and $H_e$ are related by 
\be 
\frac{\pi^2 }{30}g_* T_{\rm RH}^4 = 3 H_e^2 M_{\rm Pl}^2\left(\frac{a_e}{a_{\rm RH}}\right)^3,
\ee 
where $g_* = 106.75$ is the number of relativistic degrees of freedom in the plasma, and recall that we have taken $a_{\rm RH}/a_e = 3.34$. 

We show $\Omega h^2/0.12$ for $10^7 < H_e/{\rm GeV} < 10^{13}$. For this particular choice of $g$, the vectors can be produced in the correct abundance for $10^8 \lesssim H_e/{\rm GeV} \lesssim 10^{13}$. The solid lines correspond to our numerical results, while the dashed lines correspond to a high-$\tilde{m}$ extrapolation following $a^3 n/(a_eH_e)^3 \propto e^{-2.8\tilde{m}}$. For a given value of $H_e$, it is possible to produce both low mass and high mass particles in the correct abundance. For example, for $H_e = 10^{12}$ GeV, it is possible to obtain the correct relic abundance for both $m \sim 10$ GeV and $m \sim 10^{13}$ GeV, obtaining the latter from extrapolating to higher masses. For large values of $H_e \sim 10^{13}$ GeV, it is possible to produce dark photon DM at the GeV scale. Figure~\ref{fig:DM} is of course just one example of the viable parameter space. From a modest increase in the coupling, as in Figure~\ref{fig:varyg}, the value of $a^3 n /a_e^3H_e^3$ can increase by several orders of magnitude, further widening the parameter space. Different choices of $\tilde{a}_{\rm RH}$ will additionally raise or lower the final particle abundance and therefore allow for a larger range of possibilities in $\{ H_e, m\}$ as well.

\section{Conclusions}
\label{sec:conclude}
In this paper we have discussed the gravitational production of massive dark photons with a chiral coupling to the inflaton, and the two mechanisms that can substantially enhance the abundance of dark photons as a result. The first mechanism is an extended tachyonic phase prior to the end of inflation for one of the $A_\pm$ helicity modes, that leads to preferential production of the tachyonic mode and an overall chiral asymmetry. The second mechanism enhancing particle production arises from the post-inflationary oscillations of the inflaton, which drives a Mathieu-like resonant enhancement of both $A_\pm$ modes. The ensuing particle number density is maximized and nearly mass-invariant for $\tilde m \lesssim g$, then drops exponentially for large $\tilde m$. We characterized the particle production in both late and early reheating scenarios, showing that the total particle number is well defined in early reheating and information about the time of reheating is encoded in the dark photon number density spectra. In late reheating, enhanced growth continues as long as the inflaton oscillates, leading to a problematic runaway effect. Finally, we found the viable dark matter parameter space for these massive vectors, finding that the chiral dark photons can account for the dark matter for a wide range of $\{H_e, \tilde m\}$. 

There are a number of directions for future work. In our analysis, we have limited ourselves to small values of the coupling, $g \ll 1$, such that we can safely neglect any effects due to backreaction of the dark photons onto the dynamics of the inflaton. It would be useful to perform a lattice calculation, as in \cite{Adshead:2015pva} to fully understand the consequences for large coupling when backreaction becomes relevant. For the larger values of $g$ there will likely be significant further enhancement in the number density, which consequently widens the viable dark matter parameter space. 

A further observational signature of the GPP of massive, chiral, vectors is a correlated gravitational wave signature. In the massless case, the violent production of gauge fields is expected to source gravitational waves, as discussed in e.g., \cite{Adshead:2018doq, Dimastrogiovanni:2016fuu} leading to a circularly polarized GW background. We expect that similar results will apply in our scenario. Assuming that the chiral vectors are the dark matter, a corresponding GW signal would be a clean, independent prediction that could be observed current or future gravitational-wave detectors \cite{Cook:2011hg}. Similarly, if dark matter is indeed chiral, there are observational consequences for the formation of solitons and halos, that can arise as a signature of our model \cite{Jain:2021pnk, Alexander:2026fam}. Finally, it would be interesting to probe how our results impact the prospects for baryogenesis from massive vectors. Overall, the production of massive, chiral vector fields during inflation provides a well motivated origin for dark matter, with a multitude of potential observational signatures, warranting future exploration.

\acknowledgments
LJ is supported by the Provost's Postdoctoral Fellowship at Johns Hopkins University. This work was supported at JHU by NSF Grant No.\ 2412361, NASA ATP Grant No.\ 80NSSC24K1226, and the Templeton Foundation. 


\bibliography{ref}

\end{document}